\documentclass[apl,preprint,aps,amsmath,amssymb,showpacs,superscriptaddress]{revtex4-1}%

\usepackage{amssymb}
\usepackage{graphicx}
\usepackage{epsfig,amsmath}

\newcommand{\f}{$F \kern 2pt$}
\newcommand{\rf}{$\mathfrak{F} \kern 2pt$}
\newcommand{\s}{$G_d \kern 2pt$}
\newcommand{\vv}{$V_{ds} \kern 2pt$}
\newcommand{\n}{$S_{I}(V_{ds}) \kern 2pt$}
\newcommand{\te}{$T_{e} \kern 2pt$}

\begin{document}

\title{Electron-phonon coupling in single walled carbon nanotubes determined by shot noise}

\author{F. Wu}
\author{P. Virtanen}
\affiliation{Low Temperature Laboratory, Aalto University, Espoo, Finland}
\author{S. Andresen}
\affiliation{Niels Bohr Institute, University of Copenhagen, 2100 K\o benhavn \O, Denmark}
\author{B. Placais}
\affiliation{Ecole Normale Sup\'erieure, Laboratoire Pierre Aigrain, 24 rue Lhomond, 75005 Paris, France}
\author{P.J. Hakonen}
\affiliation{Low Temperature Laboratory, Aalto University, Espoo, Finland}

\date{\today} 

\begin{abstract}

We have measured shot noise in metallic single-walled carbon
nanotubes of length $L=1$ $\mu$m and have found strong suppression of noise with
increasing voltage.  We conclude that the coupling of electron and phonon baths at temperatures $T_e$ and $T_{ph}$  is described at
intermediate bias (20 mV $<$ \vv $\lesssim $ 200 mV)
by heat flow equation $P=\Sigma L (T_e^3-T_{ph}^3)$
where $\Sigma \sim 3 \cdot 10^{-9}$ W/mK$^3$
due to electron interaction with acoustic phonons, while at higher voltages optical phonon - electron interaction leads
to $P =\kappa _{op} L [ N (T_e)-N(T_{ph})]$ where $N(T)=
1/(\exp(\hbar\Omega/k_BT)-1)$
with optical phonons energy $\hbar \Omega$ and
$\kappa _{op}=2 \cdot 10^{2}$ W/m.

\end{abstract}

 \bigskip

\maketitle

Theoretical models based on mean-free path type of arguments
have successfully been employed to explain experimental current-voltage characteristics
of single-walled carbon nanotubes (SWNTs), and they indicate optical phonon generation with high phonon
temperatures in measurements at large bias voltages
\cite{Yao2000,Pop2005,Lazzeri2005,Lazzeri2006,Pop2007}.
In addition, time-resolved photoelectron spectroscopy has been used  to probe
energy relaxation between electrons and phonons in carbon nanotubes
\cite{Hertel2002,Moos2003}.
Recently, high-bias electron transport studies in conjunction with Raman spectroscopy
have been performed and direct confirmation of the high phonon
temperatures of several hundred Kelvin has been obtained
\cite{Bushmaker2007,Oron-Carl2008}.

These investigations have
addressed only the phonon temperature and the electronic temperature
has not been determined. We have studied shot noise in single-walled
nanotubes at high bias and employed the noise
to determine the electronic temperature. Assuming that acoustic phonons remain
at the substrate temperature, we can determine the relation for the heat flux
between electron and phonon baths in SWNTs \cite{Allen1987}.
We compare our results with those obtained by Raman spectroscopy and find close agreement with
the reported optical phonon temperatures.


For large electron-phonon (e--ph) or electron-electron scattering rates, the
solutions of the diffusive Boltzmann equation tend towards a Fermi
function, i.e., to a local equilibrium \cite{Arai1983}:
$f(\varepsilon ,{\rm x})\approx f_{0} (\varepsilon ,V({\rm x}),
T({\rm x}))\equiv \frac{1}{e^{(\varepsilon -V({\rm x}))/T({\rm x})}
+1}$, characterized by a local potential  $V({\rm x})$  and a
temperature  $T({\rm x})$. Considering a 1D wire this yields for the Fano factor
\begin{equation}\label{noiseAStemp}
   \begin{array}{l} \mathfrak{F} \equiv \frac{S_I}{2eI} =\frac{2k_B}{LeV} \int _{0}^{L} {\rm d}x\int _{-\infty}^{\infty} {\rm d}\varepsilon f(\varepsilon ,x)[1-f(\varepsilon ,x)] \\ \quad \quad \quad =\frac{2k_B}{LeV} \int _{0}^{L} {\rm d} x T(x) \equiv \frac{2 k_B T_{e}}{eV}, \end{array}
\end{equation}
where $S_I$ denotes the shot noise power and $T_{e} $  is the average electronic temperature. If thermal
conduction is dominated by electronic conduction, the Boltzmann
equation yields with the phonon bath at $T_{ph}=0$: $T_{e} =\frac{\sqrt{3}}{8} \frac{eV}{k_B}$ and
$\mathfrak{F}=\frac{\sqrt{3}}{4}$, the well-known
theoretical estimate due to hot electrons at an internal equilibrium
\cite{Nagaev1995}.


Besides the electronic heat conduction $P_{{\rm diff}} = \frac{\pi^2}{3} \frac{k_B^2}{e^2} T(x) \frac{dT(x)}{dx}$,
power flow $P_{{\rm inel}} $  between electrons and phonons has to
be taken into account when determining $T(x)$ \cite{Tarkiainen2003a}. We have
considered the standard energy balance model for the electron-sample
phonon-substrate coupling, described for example in Fig. 1 of Ref.
\onlinecite{Wellstood1994}, in which the Joule heating $P_{{\rm Joule}} $ dissipates either to the diffusive
reservoir $P_{{\rm diff}} $ or to the
lattice via inelastic scattering $P_{{\rm
inel}} $.
In such a model, the relative magnitude of  $P_{{\rm diff}} $  to  $P_{{\rm
inel}} $  determines the magnitude of the noise:  $(eV)^2-\frac{64}{3}(k_B T_e)^2=\frac{1}{\epsilon_{{\rm Th}}} P_{{\rm inel}}(T_e)$, valid in the limit $T_{ph} \rightarrow 0$; here $\epsilon_{{\rm Th}}=$ denotes Thouless energy.
In Ref. \cite{pauli} the configuration for a typical
nanotube sample is analyzed. Its conclusion is that, at intermediate voltages, the
electron-phonon heat transfer is the bottleneck and, consequently, shot noise can be
employed to obtain information on e-ph coupling.

For the dissipated power $P_{{\rm inel}}$
via e-ph interaction with Debye-like acoustic phonon
spectrum \cite{Bergmann1990}, one obtains when Debye temperature $
\theta_D = \hbar \omega _{D}/k_B \gg T$ (see also \cite{Allen1987}):
\begin{equation}\label{powerAcoustic}
  P_{{\rm inel}} = \Sigma L [ T^{\alpha+3} - T_{ph}^{\alpha+3}] \\
\end{equation}
where $\Sigma$ specifies the strength of the e-ph interaction per unit length,
and
the exponent $\alpha =0$ for a 1D sample. In general, $\alpha$ depends on the
dimensionality of the electron and phonon systems, disorder and
possibly on other factors \cite{Sergeev2005,Karvonen2005}.


For a single band of optical phonons with energy $\hbar\Omega $,
$P_{{\rm inel}} $  is given by
\begin{equation}\label{powerOptical}
P_{{\rm inel}} =\kappa _{op}  [\coth
(\frac{\hbar \Omega}{2k_B T} )-\coth (\frac{\hbar \Omega}{2k_B T_{ph}} )]{\kern 1pt} ,
\end{equation}
where  $\kappa_{op} $  describes the strength of the interaction
between electrons and acoustic phonons via optical phonon modes \cite{Tse2009}: $ \kappa_{op} = \kappa_{e-op} \kappa_{op-ac}/(\kappa_{e-op} + \kappa_{op-ac})$, where $\kappa_{e-op}$ denotes the coupling between electrons and optical phonons and $\kappa_{op-ac}$ governs the relaxation of optical phonon branches to acoustic phonons.

In voltage-biased nanotubes, the energy of electrons $\varepsilon $ is supplied by
the voltage  $V$.
From the Debye-like acoustic phonon scattering, one
obtains $ \mathfrak{F} \propto  V^{-(\alpha +1)/(\alpha +3)},$ which yields  $\mathfrak{F}
\propto  1/V^{1/3}$ for a 1D conductor.
For the e-op scattering, the behavior at large
voltages can be approximated by the estimate
\begin{equation}\label{eq:Fopt}
\mathfrak{F}=\frac{2(\hbar \Omega /eV)}{\ln [1+ \kappa_{op} /IV ]}.
\end{equation}
where we consider e-op coupling as the major relaxation channel.


Our nanotube samples were grown with chemical vapor deposition (CVD). They were
manufactured on top of insulating sapphire substrates in order to
minimize parasitic capacitance and to reduce RF losses. Pairs of
25/15 nm Ti/Au contacts, 0.3 $\mu$m apart, were patterned between the
catalyst islands by electron beam lithography. A central top-gate,
0.1 $\mu$m wide, was deposited between the contacts. It consisted of
an insulating barrier, formed by five 2 nm Al layers, each oxidized
for 2 min at dry atmospheric conditions, followed by a 25-nm layer
of Ti for the lead itself. The tube diameters were around 2 nm.


In our measurement setup at frequency \textit{f} = 600--950 MHz, we use a
liquid-helium-cooled low-noise amplifier (LNA) \cite{Roschier2004}.
We determine the differential Fano factor  $F_{d} =\frac{1}{2e}
\frac{dS_I}{dI} $ using lock-in techniques, and obtain the average, excess noise Fano factor
by $F=\frac{1}{I} \int _{o}^{I}F_{d} dI =(S_I(I)-S(0))/(2eI)$  \cite{danneau2008}. The
nonlinearity of the IV curve of the SWNT is taken into account using the
scheme described in Ref. \onlinecite{Wu2007}. Our measured $F$ is an approximation
for the true Fano factor $\mathfrak{F}$ because, with substantial $V_{ds}$-induced sample heating, the noise does not fully cross over to the shot noise regime: the correction factor
is at most $\approx \coth \frac{2eV}{4k_BT}=\coth \frac{1}{\mathfrak{F}} \simeq 1$  within 5\% when $F<0.5$,
the main region of interest in our analysis.

\begin{figure}
\begin{center}
\includegraphics[width=0.77\linewidth]{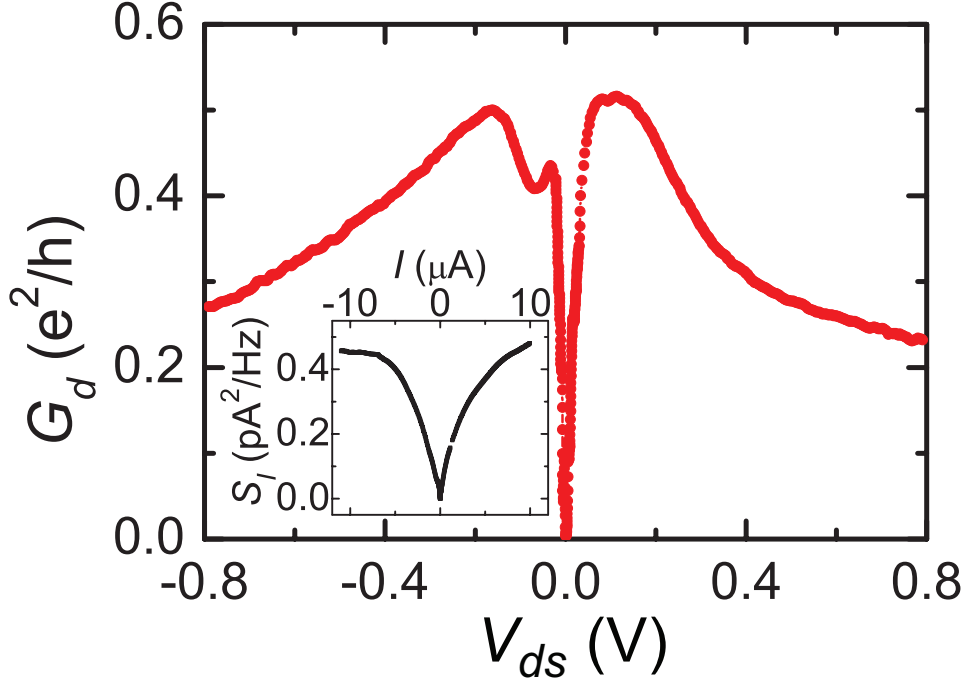}
\caption{(Color online) Differential conductance \s of a typical
SWNT nanotube sample recorded at $T = 4.2$ K. Measured power spectral density of current noise $S_I(I)$ is shown in inset.}
\label{fig:sigma}
\end{center}
\end{figure}

Fig. \ref{fig:sigma} displays the differential conductance
$G_d=\frac{dI}{dV_{ds}}$ vs. bias voltage \vv measured at $T_0=4.2$
K. Initially, there is rather strong Coulomb blockade that
suppresses the conductance below a few millivolts.
Above the Coulomb blockade \vv $>$ 10 mV , \s increases gradually
and reaches a maximum around \vv{\kern 1pt} = 0.1-0.2 V, above which
\s starts to decrease, in a manner similar to that found by Yao et
al. \cite{Yao2000}. As in Ref. \onlinecite{Yao2000}, we model the
decrease by generation of optical phonons. Moreover, this
decrease in \s suggests that the electrical contacts on our sample
are reasonably good, since otherwise the decrease of conductance due
to optical phonon scattering could not be observed according to Ref.
\onlinecite{Yao2000}. Our maximum conductance of $\sim 0.5$ $e^2/h$
(30\% less than in Ref. \onlinecite{Yao2000}) implies a mean free
path of $\l_e \sim 60$ nm.
There is slight asymmetry in \s data in
Fig. \ref{fig:sigma}, presumably due to universal conductance
fluctuation (UCF) type of behavior.

The results of shot noise measurements
$S_I(I)$ is illustrated in the inset of Fig. \ref{fig:sigma}.
Using the Khlus formula  \cite{Blanter2000}, we may
fit to the data and find $\mathfrak{F}$ = 2 at very small voltage
(\vv $<$ 5 mV).
Large $\mathfrak{F}$ at low bias is a sign of cotunneling phenomena
which are known to enhance shot noise in SWNTs \cite{Onac2006}.
At large bias, \vv $>$ 0.5 V (above 7 $\mu$A in the inset of Fig. \ref{fig:sigma}), $S_I$ tends to saturate,
especially at negative bias voltages. This is similar to the behavior
observed in semiconducting SWNTs \cite{Chaste2010}.
\begin{figure}
\begin{center}
\includegraphics[width=0.80\linewidth]{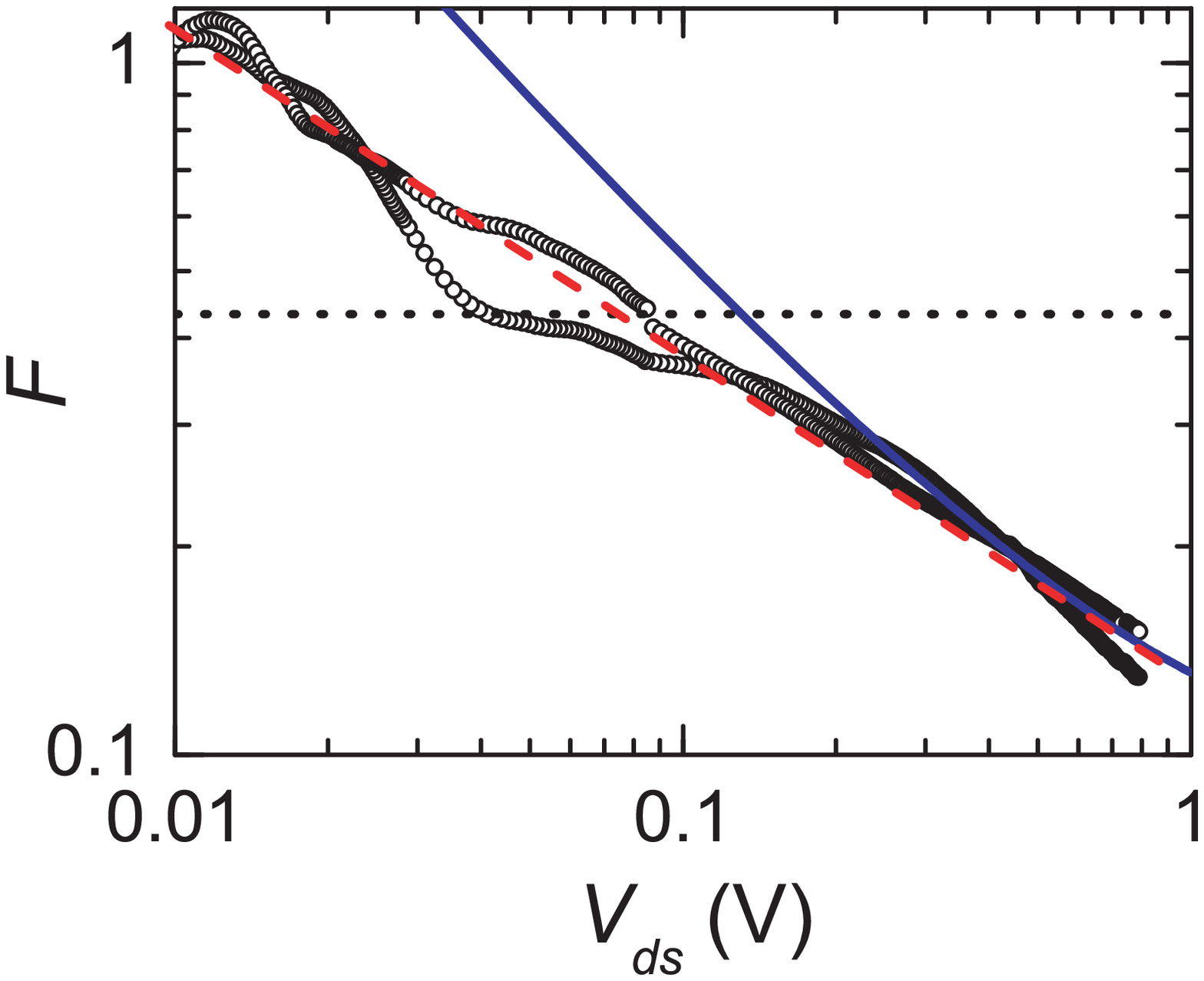}
\caption{(Color online) Fano factor \f vs.
absolute value of \vv on log-log scale. $F\propto V_{ds}^{-0.45}$
is illustrated in the figure by the dashed line. The solid blue curve
displays the fit of Eq. \ref{eq:Fopt} to the data. The dotted line denotes the hot electron value for $F$.}
\label{fig:fano}
\end{center}
\end{figure}

Fig. \ref{fig:fano} displays the Fano factor $F$ vs. \vv.
Above the cotunneling maximum in $F$ at \vv $\sim 5$ mV (not shown), the Fano factor starts to decrease.
The heat transfer
is initially dominated by diffusion along the tube and the hot
electron regime is approached. However, the hot-electron value
$\mathfrak{F}=\frac{\sqrt{3}}{4}$ is not favored in Fig. \ref{fig:fano}.
Consequently, we conclude that the noise decreases already at intermediate voltage 20 mV $<$ \vv $\lesssim$
 200 mV  due to inelastic processes. This finding signifies a relatively large inelastic scattering rate, which may be an
  indication of
  coupling to substrate modes \cite{Steiner2009}.
At higher
bias, power starts to flow out from the electronic system via
electron-optical phonon coupling, and  \f is decreased even stronger. By fitting
to the data in Fig.  \ref{fig:fano} we find  $F\propto V_{ds}^{-0.6}$
at $V_{ds}$ $>$  100 mV and $F\propto V_{ds}^{-0.45}$ at $V_{ds}$
$<$  100 mV. Other samples yielded similar values at high bias whereas somewhat larger variation in the exponent was observed at $V_{ds}$
$<$  100 mV. The
semiclassical model with acoustic phonon scattering is thus only qualitatively consistent with our data.
However, optical phonon scattering
described by Eq. \ref{eq:Fopt} is found to agree well with the high
bias data  at \vv $\gtrsim$ 0.2 V using $\kappa _{op}=2 \cdot 10^{2}$ W/m).

\begin{figure}
\begin{center}
\includegraphics[width=0.80\linewidth]{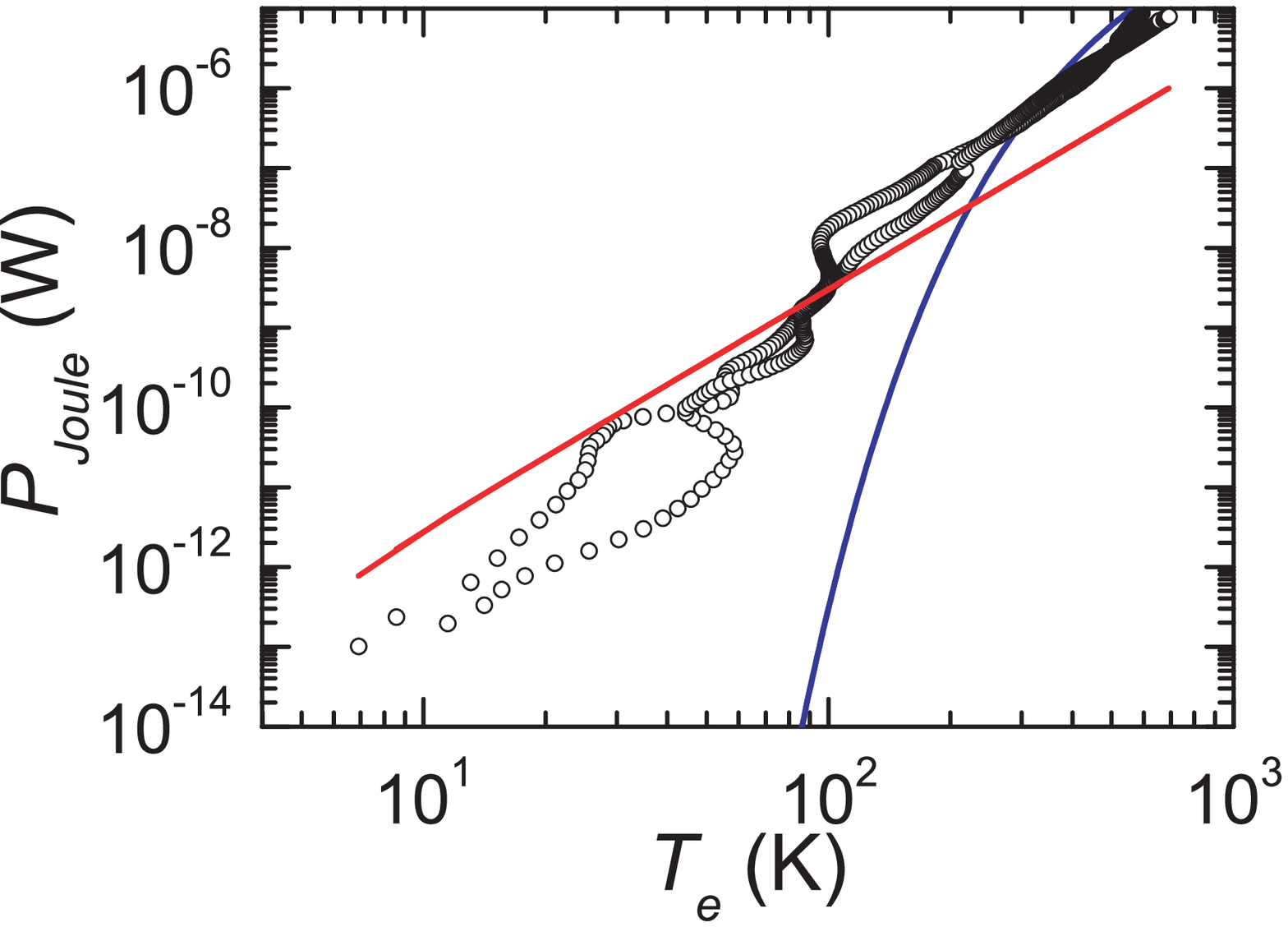}
\caption{(Color online) Joule heating $P_{\rm{Joule}}=IV_{ds}$ vs. electronic temperature \te $=\frac{F eV_{ds}}{2k_B}$.
Theory curves for acoustic phonons from Eq. \ref{powerAcoustic} ($\Sigma \sim 3 \cdot 10^{-9}$ W/mK$^3$)  and for optical phonons from Eq. \ref{powerOptical} ($\kappa _{op}=2 \cdot 10^{2}$ W/m)
are illustrated by the red and blue solid curves, respectively. }
\label{fig:power}
\end{center}
\end{figure}
  Encouraged by the success of semiclassical modeling,
  we have employed our shot noise results for thermometry
  to determine the average electronic temperature \te on the sample according to Eq. \ref{noiseAStemp}.
  Figure \ref{fig:power} displays the total heat flow due to dissipated power $P_{\rm{Joule}}=I V_{ds}$ vs. \te deduced from \f. In order to estimate $\Sigma$ and $\kappa_{op}$, we neglect $P_{{\rm diff}}$, the contribution of which is small at high bias.
  A fit using $T_e^4 - T_{ph}^4$, with $T_{ph}=T_0$, would
work the best over the whole range of data, consistent with an exponent
of $\alpha=1$ in the Debye-like spectrum. 
This would correspond to graphene-like 2-dim behavior \cite{Viljas2010} which
could point towards strongly modified phonon modes by the presence of the SiO$_2$ substrate \cite{Steiner2009}.

 At the intermediate bias range 20 mV $<$ \vv $\lesssim$
 200 mV (\te $\simeq$ 100 K), we are also able to fit
 the exponent $\alpha=0$ with the data as seen by the
 red solid curve in Fig. \ref{fig:power}. This
 yields $P_{inel}=\Sigma L(T^3-T_0^3)$ with $\Sigma \sim 3 \cdot 10^{-9}$ W/mK$^3$.
 At \vv $\gtrsim 0.2$ V (\te $\gtrsim$ 350 K), optical phonons take
 over and we obtain a good fit of Eq. \ref{powerOptical} to the
 data using $\kappa _{op}=2 \cdot 10^{2}$ W/m and
 $\hbar \Omega=0.18$ eV.
 \cite{note}.
Our result displays a different power law compared with the work of
Moos et al. \cite{Moos2003} who obtain a relation of $T_e^5-T_0^5$ for a
nanotube bundle. Our low-bias dependence $P \propto T^3$ agrees with
the result of Appenzeller \emph{et al}. who report temperature dependence $\propto 1/T_e$ for the
electron-phonon scattering time  \cite{Appenzeller2001}.

In conclusion, using diffusive transport theory and shot noise measurements in SWNTs at high bias, we determined
 the electronic temperature which nearly coincides with phonon temperatures
obtained recently by Raman spectroscopy in Refs.
\onlinecite{Oron-Carl2008,Deshpande2009,Steiner2009}. Consequently, optical
phonons and electrons are nearly at the same temperature, which is in agreement
with standard heat flow modeling with typical electron-phonon coupling parameters \cite{Lazzeri2006}.

We wish to thank V. Ermelov, T. Heikkil\"a, F. Mauri, N. Vandecasteele, and J. Viljas for useful discussions and correspondence. This work was supported by the Academy
of Finland (Materials World Network), EU FP6-IST-021285-2, and the NANOSYSTEMS project with Nokia Research Center.


\begin{thebibliography}{1}
\bibitem{note}%
  \BibitemOpen
  \bibinfo {note} {Our shot noise results are not affected by contact
  resistance because if it were significant, then Fano factor would remain
  constant and the $T$-dependence in Fig. \ref {fig:power} would be like
  $P_{{\protect \rm inel}} \propto T_e^2$, which does not agree with the
  observed behavior.}\BibitemShut {Stop}%
\end{thebibliography}

%

\end{document}